\documentclass{article}
\usepackage{graphicx}
\usepackage{url}
\usepackage{amsmath}
\usepackage{amssymb}
\usepackage{amsthm}
\usepackage{hyperref}
\usepackage{stmaryrd}
\usepackage{todonotes}

\usepackage{tikz}
\usetikzlibrary{arrows.meta}

\newtheorem{definition}{Definition}
\newtheorem{theorem}{Theorem}

\newtheorem{lemma}{Lemma}

\theoremstyle{remark}
\newtheorem{example}{Example}

\DeclareMathOperator{\Int}{Int}
\DeclareMathOperator{\Root}{root}
\DeclareMathOperator{\rep}{rep}
\DeclareMathOperator{\out}{out}
\DeclareMathOperator{\OPM}{OPM}
\DeclareMathOperator{\Tree}{Tree}

\newcommand{\lp}{\llparenthesis}
\newcommand{\rp}{\rrparenthesis}
\newcommand{\enc}{\eta}

\title{Operator Precedence Tree Automata}
\author{Matteo Pradella\\
  \small DEIB, Politecnico di Milano\\
  \small Milano, Italy\\
  \small \texttt{matteo.pradella@polimi.it}}
\date{}

\begin{document}

\maketitle
\begin{abstract}
Operator-precedence tree automata (OPTA) separate the construction
of a word's syntactic structure from its recognition: an
operator-precedence (OP) matrix determines a canonical tree, which
is evaluated by a finite bottom-up automaton with regular
horizontal transition languages. 
OPTA define exactly the operator-precedence languages. 
For a fixed matrix and language, a tree-context
congruence yields an effectively constructible canonical
complete deterministic exact recognizer, minimal in the number
of vertical states and unique up to isomorphism. 
For acyclic OP matrices,
we characterize balanced
non-counting by aperiodicity of the context transition monoid
of the canonical minimal ranked tree recognizer. 

\end{abstract}

\section{Introduction}
\label{sec:introduction}

Operator-precedence languages (OPLs) originated in Floyd's work
on deterministic parsing of programming languages~\cite{Floyd1963}.
Their defining feature is a finite matrix of precedence relations
between terminal symbols, which determines how a word is
decomposed into nested syntactic units. Beyond their original
application, OPLs form a family of deterministic
context-free languages, properly containing \emph{Input Driven}
or \emph{Visibly Pushdown}
languages~\cite{DBLP:conf/icalp/Mehlhorn80,jacm/AlurM09} and retaining 
many of the properties that make regular
languages useful. These include Boolean and concatenation
 closure under compatible
precedence assumptions \cite{Crespi-ReghizziM12}, and
automata-theoretic and logical characterizations
\cite{LonatiEtAl2015,MP18}. Their local parsability also supports
the analysis of separate input fragments, providing a basis for
parallel parsing~\cite{BarenghiEtAl2015,CMP25}.

Tree automata offer another way of extending finite-state methods
beyond words. They recognize explicitly structured objects by
assigning states to tree nodes according to the states computed
on their children. Their theory provides context equivalences,
determinization and minimization constructions, and connections
with grammars and logic~\cite{TATA}. For unranked trees,
regular languages describe the admissible sequences of child
states, allowing finite descriptions even when node arity is
unbounded~\cite{CristauLoedingThomas2005}.
These features suggest a natural connection with operator
precedence: the matrix supplies a tree structure that a finite
tree automaton can subsequently evaluate.

This connection brings together three views of syntax.
A grammar gives a generative description of a language and its
derivations. A matrix-driven OP parser gives an operational
description: for a fixed precedence matrix $M$, it already acts
as a recognizer of the compatible-word language $L(M)$ and
determines a canonical tree $T_M(w)$ for each accepted word.
A tree automaton, in turn, takes a tree as input and decides
whether that supplied structure belongs to its language.
In an operator-precedence automaton (OPA), structural parsing
and the finite-state control selecting a particular language
are performed together. Making their respective roles explicit
allows the structural component of OP parsing to be combined
directly with the recognition mechanisms of tree automata.

We introduce \emph{operator-precedence tree automata} (OPTA)
to formalize this combination. An OPTA consists of an OPM and
a finite tree recognizer whose horizontal transition languages
are regular. The matrix determines the OP tree of the input
word; the recognizer then evaluates that tree from the leaves
to the root. Terminal leaves retain their letters, while all
internal nodes carry the same structural label, so neither
grammar nonterminals nor boundary annotations need to be supplied
as part of the input tree. This separates the construction of
the structure from the acceptance condition and makes it possible
to study the same language both as a set of words and as a set
of canonical trees.

The connection also exposes an essential difference between the
two theories. Concatenation of OPLs is ordinary word
concatenation, whereas the products underlying regular tree
expressions are based on substitution at designated
leaves~\cite{TATA}. Similarly, composition of one-hole tree
contexts substitutes one context into another. These operations
cannot be identified merely by passing from a word to its tree:
the structure assigned to a concatenation $uv$ need not preserve
the separately constructed trees of $u$ and $v$ as subtrees.
Consequently, transferring a statement about regular expressions
or star-freeness requires attention to the operation on which
the expression formalism is based.


Our results develop this connection in both directions.
First, we prove that OPTA recognize exactly the OPLs over a
fixed OP alphabet. The equivalence holds for arbitrary
conflict-free matrices, including matrices with cycles in the
equal-precedence relation; regular horizontal languages account
for the unbounded arities that may arise in this case.

Second, we apply the tree-context Myhill-Nerode construction to
obtain a canonical minimal \emph{exact} OPTA. Exactness requires
recognition of precisely the canonical tree language $T_M(L)$,
including rejection of trees outside the OP-tree universe.
This convention yields an effectively constructible minimum in
the number of vertical states, unique up to isomorphism, for
every conflict-free matrix. We also show why specifying behavior
only on valid OP trees need not yield a unique minimum.
The horizontal representation can be minimized separately once
the canonical vertical quotient has been fixed.

Third, for acyclic matrices, we characterize
balanced non-counting by aperiodicity of the context transition
monoid of the canonical minimal ranked tree recognizer.

The transfer is therefore two-way: tree-automata methods provide
canonical minimization for OP languages, while the logical and
algebraic theory of OPLs yields aperiodicity characterizations
for their canonical tree encodings. 

A closely related contribution is the Myhill-Nerode
characterization of OPLs established in~\cite{Henzinger23}. 
Their syntactic congruence is
defined on words over a fixed operator-precedence alphabet:
it refines chain equivalence and tests indistinguishability
under embeddings compatible with the chain structure.
They prove that this congruence has finite index exactly
for OPLs, and exploit it to develop antichain algorithms
for language inclusion and universality.

Our approach provides a complementary perspective by applying
the tree-context Myhill-Nerode construction to the exact
tree language $T_M(L)$.
The objects being compared are internal-rooted trees, and
the tests are substitutions into arbitrary one-hole tree
contexts, with membership in $T_M(L)$ fixing rejection of
invalid OP trees.
The resulting equivalence classes are the vertical states
of a canonical complete deterministic exact OPTA, minimal
among such recognizers and unique up to isomorphism.
This minimality statement concerns exact tree recognition
for a fixed matrix; it does not assert minimality among
ordinary operator-precedence automata.
For acyclic matrices, the ranked presentation
provides a context transition monoid whose aperiodicity
characterizes balanced non-counting.
Hence, the tree representation connects canonical minimization
with the algebraic analysis of non-counting OPLs.


The paper is structured as follows. 
Section~\ref{sec:background} recalls the required background.
Section~\ref{sec:opta} introduces OPTA and proves their equivalence
with OPLs. Section~\ref{sec:minimality} develops canonical exact
recognition and minimality, and Section~\ref{sec:aperiodicity}
establishes the aperiodicity characterizations.

\section{Background}
\label{sec:background}

We assume familiarity with finite state automata, regular languages,
context-free grammars, and tree languages \cite{Salomaa73,TATA}. 
Throughout, $\Sigma$ is a finite nonempty
alphabet, $\varepsilon$ is the empty word, and
$\Sigma^+=\Sigma^*\setminus\{\varepsilon\}$.
We study languages of nonempty words; empty factors and auxiliary
languages containing $\varepsilon$ are nevertheless allowed.
The symbol $\#$, not in $\Sigma$, delimits words, 
and $\Sigma_\#=\Sigma\cup\{\#\}$.

\subsection{Trees and tree automata}

All trees are finite, rooted, and ordered. For a tree $t$, we write
$\Root(t)$ for its root and $\Int(t)$ for its set of internal nodes,
that is, nodes having at least one child. Its \emph{yield} is the
word obtained by reading its leaf labels from left to right.
A ranked alphabet $\Gamma$ assigns an arity to each symbol;
$T_\Gamma$ denotes the trees in which every node has the arity
prescribed by its label. In an unranked tree, the label does not
determine the number of children.

A {\em bottom-up tree automaton} computes a state at a node from its
label and the states computed at its children. In the ranked case,
a deterministic complete automaton has, for each symbol $f$ of
arity $k$, a total transition function $\Delta_f:Q^k\to Q$;
for $k=0$, this specifies a state for the leaf $f$.
A tree is accepted when its root state is final.
In the unranked case, transitions can instead be specified by
regular languages of child-state sequences: a transition
$H\xrightarrow{f}q$, with $H\subseteq Q^*$ regular, permits state
$q$ at an $f$-node precisely when its child-state word belongs
to $H$~\cite{CristauLoedingThomas2005}.
Tree languages recognized by these devices are called
\emph{regular}. Bottom-up nondeterministic automata can be
determinized in both settings. A deterministic tree automaton is
\emph{reduced} if every state is the root state of some input
tree; that tree need not be accepted~\cite[Section~1.1]{TATA}.

A \emph{tree context} has one distinguished hole $\Box$ in place
of a subtree. We write $C[t]$ for substitution of $t$ into the
hole, and $C[D]$ for composition of contexts. Iteration is defined
by $C^0=\Box$ and $C^{n+1}=C[C^n]$.

\subsection{Precedence matrices and chains}

We recall here some basic definitions from the theory of 
Operator-Precedence languages \cite{Floyd1963,LonatiEtAl2015,MP18}.

A conflict-free \emph{operator-precedence matrix} (OPM) is a
partial map
\[
    M:\Sigma_\#\times\Sigma_\#
      \to \{\lessdot,\doteq,\gtrdot\}.
\]
We write $a\lessdot b$, $a\doteq b$, or $a\gtrdot b$ when
$M(a,b)$ has the corresponding value, read as \emph{yields},
\emph{is equal in}, or \emph{takes precedence}, respectively.
These relations need not have the usual properties of an order.
We use the boundary convention
\[
    \#\lessdot a,\qquad a\gtrdot\#\quad(a\in\Sigma),
    \qquad \#\doteq\#.
\]
The pair $(\Sigma,M)$ is an \emph{OP alphabet}.
Matrix inclusion is entrywise: $M'\subseteq M$ means that every
relation specified by $M'$ is also specified by $M$.
The matrix is \emph{total} if every terminal-pair entry is defined,
and \emph{acyclic} if the directed graph of
$\doteq$ on $\Sigma$ has no cycle.

Chains describe the structure induced by $M$ independently of
grammar nonterminals.
A \emph{simple chain}, written $a_0[a_1\cdots a_k]a_{k+1}$,
has $k\geq1$, $a_1,\ldots,a_k\in\Sigma$,
$a_0,a_{k+1}\in\Sigma_\#$, $M(a_0,a_{k+1})$ defined, and
\[
    a_0\lessdot a_1\doteq\cdots\doteq a_k\gtrdot a_{k+1}.
\]
A \emph{composed chain} has the form
\[
    a_0[w_0a_1w_1\cdots a_kw_k]a_{k+1},
\]
where $a_0[a_1\cdots a_k]a_{k+1}$ is a simple chain and, for
each $0\leq i\leq k$, either $w_i=\varepsilon$ or
$a_i[w_i]a_{i+1}$ is recursively a chain.
The symbols $a_0,a_{k+1}$ are its \emph{boundaries}, the word
between brackets is its \emph{body}, and $a_1\cdots a_k$ is its
\emph{skeleton}.

A word $w\in\Sigma^+$ is \emph{compatible} with $M$ if
$\#[w]\#$ is a chain. We denote the set of such words by $L(M)$.
Conflict-freeness makes their recursive chain decomposition
unique. Replacing each chain by one internal node and retaining
its skeleton terminals and nonempty subchains as ordered children
gives the OP tree used below. A skeleton of length $k$ gives at
most $2k+1$ children. If $M$ is acyclic, its skeletons
have no repeated terminal, so $k\leq|\Sigma|$; thus OP trees have
arity at most $2|\Sigma|+1$.
Cyclic matrices do not impose any arity bound.

\subsection{Grammars and operator-precedence automata}

We use grammars $G=(\Sigma,V_N,P,S)$ with a finite nonterminal
set $V_N$ disjoint from $\Sigma$, a finite production set $P$,
and a nonempty set of axioms $S\subseteq V_N$.
Empty and renaming productions are excluded. A production is in
\emph{operator form} if its right-hand side has no adjacent
nonterminals. For such grammars, put
\[
\begin{aligned}
    \mathcal L_G(A)
      &=\{a\in\Sigma\mid A\Rightarrow^*Ca\alpha,
                         \ C\in V_N\cup\{\varepsilon\}\},\\
    \mathcal R_G(A)
      &=\{a\in\Sigma\mid A\Rightarrow^*\alpha aC,
                         \ C\in V_N\cup\{\varepsilon\}\},
\end{aligned}
\]
where $\alpha$ ranges over $(\Sigma\cup V_N)^*$.
The induced matrix $\OPM(G)$ collects the relations
\[
\begin{aligned}
 a\doteq b
   &\quad\text{if } A\to\alpha aCb\beta\in P,
       \quad C\in V_N\cup\{\varepsilon\},\\
 a\lessdot b
   &\quad\text{if } A\to\alpha aB\beta\in P,
       \quad b\in\mathcal L_G(B),\\
 a\gtrdot b
   &\quad\text{if } A\to\alpha Bb\beta\in P,
       \quad a\in\mathcal R_G(B),
\end{aligned}
\]
together with the boundary relations for the terminal sets of
the axioms and $\#\doteq\#$. An operator grammar is an \emph{operator-precedence
grammar} (OPG) when $\OPM(G)$ is conflict-free~\cite{MP18}.

The extension to cyclic matrices uses \emph{cyclic OPGs}
(C-OPGs)~\cite{CMP25}.
Their right-hand sides are regular expressions built using
concatenation and positive iteration, subject to operator form;
alternative productions provide finite union.
An occurrence of $A\to r$ chooses a word $\alpha\in L(r)$ as
the child sequence of that production occurrence. The precedence
relations are computed over these instantiated right-hand sides,
and must remain conflict-free. This permits unbounded child
sequences without introducing auxiliary nonterminal nodes.

An \emph{operator-precedence automaton} (OPA) has a finite control
set $Q$, initial and final sets $I,F\subseteq Q$, a matrix $M$,
and transition functions
\[
    \delta_{\mathrm{push}},\delta_{\mathrm{shift}}:
        Q\times\Sigma\to2^Q,
    \qquad
    \delta_{\mathrm{pop}}:Q\times Q\to2^Q.
\]
Stack entries are pairs $[a,p]\in\Sigma\times Q$, above a
bottom marker $\bot$ whose terminal symbol is $\#$.
Let $q$ be the current state, $a$ the top terminal, and $b$ the
next input symbol. The matrix determines the move:
\begin{itemize}
    \item If $a\lessdot b$, consume $b$, push $[b,q]$, and
          choose a state in $\delta_{\mathrm{push}}(q,b)$.
    \item If $a\doteq b$, consume $b$, replace the top entry
          $[a,p]$ by $[b,p]$, and choose a state in
          $\delta_{\mathrm{shift}}(q,b)$.
    \item If $a\gtrdot b$, remove $[a,p]$ and choose a state
          in $\delta_{\mathrm{pop}}(q,p)$, without consuming $b$.
\end{itemize}
Shift and pop require a nonbottom entry.
Acceptance starts with stack $\bot$ and a state in $I$, 
and ends with stack $\bot$,
a state in $F$, and only the final $\#$ unread.
An undefined relation blocks the run.
A \emph{support} of a chain $a[x]b$ is the computation fragment
that reads its body $x$, with $b$ as the final lookahead, and
returns to the underlying stack prefix after closing the chain.

We call languages accepted by OPAs \emph{operator-precedence
languages} (OPLs). OPAs and OPGs define the same languages over
an acyclic matrix \cite{LonatiEtAl2015}.
C-OPGs provide the corresponding grammatical characterization
for arbitrary conflict-free matrices~\cite{CMP25}.

\subsection{Logical and algebraic background}

For $w=a_1\cdots a_n\in L(M)$, set $a_0=a_{n+1}=\#$.
The associated word structure has domain $\{0,\ldots,n+1\}$,
the linear order $<$, letter predicates $P_a$ for
$a\in\Sigma_\#$, and the \emph{chain relation}
\[
    \chi(i,j)
    \quad\Longleftrightarrow\quad
    i<j\quad\text{and}\quad
    a_i[a_{i+1}\cdots a_{j-1}]a_j\text{ is a chain}.
\]
First-order logic (FO) quantifies over positions; monadic
second-order logic (MSO) also quantifies over sets of positions.
These are logics of the structured words, rather than logics
whose domain is the nodes of their OP trees.

Precedence-Oriented Temporal Logic (POTL) is interpreted on the
same OP-word structures. It extends linear temporal reasoning
with modalities that navigate the chain structure.
POTL is expressively equivalent to FO with one free variable
on finite OP words~\cite[Theorem~4.9]{CMP21arXiv}.
We say that $L\subseteq L(M)$ is POTL-definable if there is a
POTL formula $\varphi$ such that
\[
    L=\{w\in L(M)\mid (w,0)\models\varphi\},
\]
where position $0$ carries the initial boundary marker $\#$.

Operator-precedence expressions (OPEs) extend regular expressions
with Boolean operations and a \emph{fence} operation that tests
the parenthesized structure assigned by $M$.
We use the syntax and semantics of~\cite{MPC23}, where they are proved 
to define OPLs. 
An OPE is \emph{star-free} if it uses neither Kleene star nor positive iteration.
The totality convention in that reference is without loss of
generality: undefined entries of $M$ can be enforced by its
\emph{hole} constraints after completing $M$.
These constraints admit star-free expressions, and a completion
adding only $\lessdot$ or $\gtrdot$ entries preserves
acyclicity. The same reference
establishes the FO definability of star-free OPEs,
and their non-counting property.
The precise non-counting
condition is recalled in the aperiodicity section, where it is
related to OPTA aperiodicity.

Finally, a finite monoid $V$ is \emph{aperiodic} if there exists
$N\geq1$ such that $v^N=v^{N+1}$ for every $v\in V$.
In this paper, the relevant monoids consist of transformations
induced by tree contexts.

\section{OP Tree Automata}
\label{sec:opta}

Let $\Sigma$ be a finite alphabet and let $M$ be an operator-precedence matrix over $\Sigma$. 
Write $L(M) \subseteq \Sigma^+$ for the set of words compatible with $M$.

Let $\mathcal T_\Sigma^\bullet$ be the set of all finite ordered
trees whose leaves are labeled in $\Sigma$, whose internal nodes
are labeled $\bullet$, and whose root is an internal node.
Every internal node has a positive, otherwise unrestricted,
number of children. Membership in $\mathcal T_\Sigma^\bullet$
does not require compatibility with $M$.

For $w\in L(M)$, let $T_M(w)$ denote its operator-precedence tree, obtained from the unique chain
decomposition determined by $M$. 
Each chain reduction contributes one internal node labeled $\bullet$, and the leaves carry the letters of $w$ in their original order. 
The children of a chain node are its skeleton terminals and the trees of its subchains, in left-to-right order; boundary markers are omitted.

For $L\subseteq L(M)$, write
\[
    T_M(L)=\{T_M(w)\mid w\in L\},
    \qquad U_M=T_M(L(M)).
\]
Thus $U_M$ is the universe of valid OP trees for $M$, and
\[
    T_M(L)\subseteq U_M\subseteq\mathcal T_\Sigma^\bullet.
\]

\begin{definition}[Operator-precedence tree automaton]
\label{def:opta}
An OPTA is a tuple
\[
    \mathcal A
    = \bigl(\Sigma,M,Q,F,(H_q)_{q\in Q}\bigr),
\]
where $\Sigma$ is a finite terminal alphabet, $M$ is an OPM over $\Sigma$,
$Q$ is a finite set of states, also called {\em vertical states},
$F\subseteq Q$ is the set of final states, and each
\[
    H_q \subseteq (\Sigma \uplus Q)^+,
\]
where $\uplus$ denotes disjoint union, 
is a regular language, called the 
\emph{horizontal transition language} associated with $q$.%
\footnote{Horizontal words are nonempty because internal nodes have positive
arity; allowing $\varepsilon$ in an $H_q$ would not affect any run.}

Let $t\in\mathcal T_\Sigma^\bullet$.
A \emph{run} of $\mathcal A$ on $t$ is a mapping
\(
    \rho: \Int(t) \to Q
\)
such that, for every internal node $v$ with ordered children
$v_1,\ldots,v_k$,
\[
    \lambda_\rho(v_1)\cdots\lambda_\rho(v_k)
    \in H_{\rho(v)},
\]
where
\[
    \lambda_\rho(u)=
    \begin{cases}
        a,
        & \text{if $u$ is a leaf labeled $a\in\Sigma$,}\\
        \rho(u),
        & \text{if $u$ is an internal node.}
    \end{cases}
\]
Therefore, terminal children contribute their labels, whereas
internal children contribute their computed states.

The run is \emph{accepting} if $\rho(\Root(t)) \in F$.
The unrestricted tree language $\Tree(\mathcal A)$
consists of all $t\in\mathcal T_\Sigma^\bullet$ admitting an
accepting run. The word language recognized by $\mathcal A$ is
\[
    L(\mathcal A)
    =
    \bigl\{
        w\in L(M)
        \mid
        T_M(w)\text{ admits an accepting run of }\mathcal A
    \bigr\}.
\]
\end{definition}

Horizontal words containing adjacent vertical states cannot occur
in runs on valid OP trees and therefore do not affect
$L(\mathcal A)$. They may still occur in the horizontal languages
of complete recognizers, whose runs are also defined on invalid trees.

An OPTA is deterministic when its horizontal languages are
pairwise disjoint. Deterministic and nondeterministic OPTA
recognize the same languages, by bottom-up tree-automaton
determinization~\cite{TATA}.

\begin{example}[Matching pairs with even depth]
\label{ex:opta-even-pairs}
Let $\Sigma=\{a,b\}$ and consider the language
\(
    L = \{a^{2n}b^{2n} \mid n \geq 1\}.
\)

\begin{figure}
\begin{center}
   \begin{tikzpicture}[
    level distance=8mm,
    sibling distance=12mm,
    every node/.style={inner sep=1.5pt}
]
\node[label=right:{$q_0$}] {$\bullet$}
    child {node {$a$}}
    child {node[label=right:{$q_1$}] {$\bullet$}
        [sibling distance=9mm]
        child {node {$a$}}
        child {node {$b$}}
    }
    child {node {$b$}};
\end{tikzpicture} 
\end{center}

    \caption{Accepting run on
$T_M(aabb)=\bullet(a,\bullet(a,b),b)$;
states are written beside internal nodes.}\label{fig:ex1}
\end{figure}
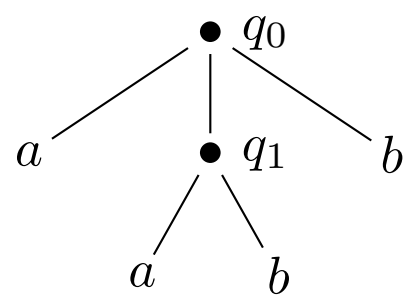

We fix the OPM $M$ such that $a \lessdot a$, $b \gtrdot b$, and
$a \doteq b$, with the usual boundary relations; all other entries are
undefined.

The following automaton accepts $L$:
\[
    \begin{gathered}
    \mathcal A=(\Sigma,M,\{q_0,q_1\},\{q_0\},
                  (H_{q_0},H_{q_1})),\\
    H_{q_0}=\{a q_1 b\},\qquad
    H_{q_1}=\{ab,a q_0 b\}.
    \end{gathered}
\]
The horizontal languages are disjoint, so $\mathcal A$ is
deterministic. Its runs alternate $q_1$ and $q_0$ along the nested
$ab$-pairs, accepting exactly when the number of pairs is even.


For instance, the accepting run on
$T_M(aabb)=\bullet(a,\bullet(a,b),b)$ is shown in Figure~\ref{fig:ex1}.
\end{example}

\subsection{OPTA Define OP Languages}

Throughout, $M$ is a conflict-free operator-precedence matrix;
no acyclicity assumption is made.
In the cyclic case, the grammatical counterpart is a cyclic
operator-precedence grammar (C-OPG), whose production right-hand
sides are regular expressions, in the sense of \cite{CMP25}.

\begin{theorem}[From cyclic OPGs to OPTA]
\label{thm:copg-to-opta}
Let $G=(\Sigma,V_N,P,S)$ be a C-OPG 
with $\OPM(G) \subseteq M$.
There is an effectively constructible OPTA $\mathcal A_G$ over
$(\Sigma,M)$ such that
\[
    L(\mathcal A_G)=L(G).
\]
\end{theorem}

\begin{proof}
For every nonterminal $A \in V_N$, let
\[
    H_A=\bigcup_{A\to r\in P}L(r),
\]
where $L(r) \subseteq (\Sigma \uplus V_N)^+$ is the regular language
denoted by the production expression $r$.
Define
\[
    \mathcal A_G
    =\bigl(\Sigma,M,V_N,S,(H_A)_{A\in V_N}\bigr).
\]
Each $H_A$ is regular.

For $w\in L(G)$, choose a derivation tree in which each production
occurrence $A\to r$ is instantiated by some word in $L(r)$.
Replacing its internal labels by $\bullet$ yields $T_M(w)$,
because $\OPM(G)\subseteq M$.
The nonterminal labels give an accepting run of $\mathcal A_G$.

Conversely, an accepting run on $T_M(w)$ assigns a nonterminal
$A$ to each internal node whose sequence of terminal children and
states of internal children belongs to $H_A$.
Choose a production $A\to r$ witnessing this membership.
These choices yield a derivation tree of $w$ in $G$.
\end{proof}

\begin{theorem}[From OPTA to OPA]
\label{thm:opta-to-opa-general}
For every OPTA
\[
    \mathcal A=\bigl(\Sigma,M,Q,F,(H_q)_{q\in Q}\bigr)
\]
there is an effectively constructible OPA $\mathcal B$ using
the same matrix $M$ such that
\[
    L(\mathcal B)=L(\mathcal A).
\]
In particular, this holds when $M$ has $\doteq$-cycles.
\end{theorem}

\begin{proof}
Since the languages $H_q$ are regular, their complete deterministic
recognizers can be combined into one finite deterministic device
with state set $S$, initial state $s_0$, transition function
\[
    \bar\delta:S\times(\Sigma \uplus Q) \to S,
\]
and output function
\[
    \operatorname{out}:S \to 2^Q
\]
such that, for every $u \in (\Sigma \uplus Q)^*$,
\[
    \out\bigl(\bar\delta^*(s_0,u)\bigr)
    =\{q\in Q\mid u\in H_q\}.
\]

The OPA $\mathcal B$ has control-state set
\[
    Q_{\mathcal B} = S \times (Q\cup\{\downarrow\}),
\]
where $\downarrow \ \notin Q$ denotes the absence of a pending subtree
state.
In a state $(s,x)$, the first component summarizes the horizontal
word processed so far, while $x\in Q$, if present, is the state
of a completed subtree not yet incorporated into that word.
For convenience, extend $\bar\delta$ by
\[
    \bar\delta(s,\downarrow)=s.
\]

We use the standard OPA convention: a push stores the current
control state on the stack, a shift preserves this saved state,
and a pop transition receives the current state and the saved
state of the removed stack entry.
The initial state of $\mathcal B$ is $(s_0,\downarrow)$,
while the set of final states is $\{(s_0,q)\mid q\in F\}$.

For $a\in\Sigma$, define
\begin{align*}
    \delta_{\mathrm{push}}\bigl((s,x),a\bigr)
      &=
      \bigl\{(\bar\delta(\bar\delta(s_0,x),a),\downarrow)\bigr\},
      \\
    \delta_{\mathrm{shift}}\bigl((s,x),a\bigr)
      &=
      \bigl\{(\bar\delta(\bar\delta(s,x),a),\downarrow)\bigr\},
      \\
    \delta_{\mathrm{pop}}\bigl((s,x),(t,y)\bigr)
      &=
      \bigl\{(t,q)\mid
          q \in \out(\bar\delta(s,x))\bigr\}.
\end{align*}
As usual, $M$ determines whether a push, shift, or pop is performed.

A push starts a new horizontal computation from $s_0$.
If a subtree state is pending, it is read first, as the leftmost
child of the new chain node.
A shift extends the current horizontal computation, first reading
any pending subtree state and then the terminal.
A pop reads any pending rightmost subtree state, chooses the
state of the completed chain node, and restores the suspended
horizontal state $t$.
The saved component $y$ is not restored: when present, it was
already incorporated as the leftmost child at the matching push.

To prove correctness, we establish the following invariant by
induction on chain trees.
For any chain with tree $t$, and any $s\in S$, its support in
$\mathcal B$, starting in control state $(s,\downarrow)$, can finish in
$(s,q)$ if and only if $t$ admits a run of $\mathcal A$ with root
state $q$.
The support consumes the chain body and returns to the same
underlying stack prefix.

Write the chain body as
\[
    w_0a_1w_1\cdots a_kw_k,
    \qquad k\geq 1.
\]
For each nonempty subchain $w_i$, let $q_i$ be its chosen root
state, and put
\[
    z_i=
    \begin{cases}
        q_i, & w_i\neq\varepsilon,\\
        \varepsilon, & w_i=\varepsilon.
    \end{cases}
\]
By the induction hypothesis, processing a nonempty $w_i$
preserves the surrounding horizontal state and leaves $q_i$
pending. The push on $a_1$, the subsequent shifts, and the final
pop therefore feed the horizontal device exactly the word
\[
    u=z_0a_1z_1\cdots a_kz_k.
\]
The pop can select exactly the states
$
    q \in \out\bigl(\bar\delta^*(s_0,u)\bigr),
$
i.e., those satisfying $u\in H_q$.
These are precisely the possible root states of an OPTA run
on $t$. The pop restores the first component $s$ saved by the
matching push. This proves the invariant in both directions;
the case with no subchains is its base case.

Applying the invariant to the outer chain $\# [w] \#$ shows that
$\mathcal B$ finishes with empty stack in $(s_0,q)$ exactly when
$T_M(w)$ has a run with root state $q$.
Both devices use the same matrix, so words outside
$L(M)$ cannot be accepted.
The set of final states of $\mathcal B$ gives
$L(\mathcal B)=L(\mathcal A)$.
\end{proof}

Hence, OPTA have the same expressive power as OPAs and C-OPGs,
defining the operator-precedence languages over arbitrary
conflict-free matrices, including cyclic ones.

\section{Canonical recognition and minimality}
\label{sec:minimality}

Let $L$ be an OPTA language on $(\Sigma, M)$ 
with no acyclicity assumption for $M$.
We use the unranked universe $\mathcal T_\Sigma^\bullet$ and
its subset $U_M$ of valid OP trees.
In this setting, a context has exactly one hole for a tree of
$\mathcal T_\Sigma^\bullet$; the identity context $\Box$ is included.
We call $\mathcal C_\Sigma^\bullet$ the set of these contexts.

\subsection{Exact recognition}
\label{sec:exact-recognition}

By Definition~\ref{def:opta}, $\Tree(\mathcal A)$
is the unrestricted tree language of $\mathcal A$.
Recognition of the word language $L$ therefore requires only
\[
    \Tree(\mathcal A)\cap U_M=T_M(L).
\]

\begin{definition}
\label{def:opta-exact-recognition}
An OPTA is an \emph{exact recognizer} of $(M,L)$ if
\[
    \Tree(\mathcal A)=T_M(L).
\]
It therefore rejects every tree outside $U_M$.

Throughout the minimality results, a normalized complete deterministic OPTA has
pairwise disjoint horizontal languages satisfying
\[
    H_q\subseteq(\Sigma\uplus Q)^+,
    \qquad
    \bigcup_{q\in Q}H_q=(\Sigma\uplus Q)^+.
\]
The positive-arity normalization is already part of
Definition~\ref{def:opta}.
Its size for the minimality statements is $|Q|$, the number
of \emph{vertical states}; terminal labels and states of auxiliary
horizontal word automata are not counted in $Q$.
\end{definition}

The exactness requirement fixes the behavior on invalid trees.
The following example shows why this matters for uniqueness.

\begin{example}[Recognition restricted to valid inputs]
\label{ex:opta-promise-minimum}
Let $\Sigma=\{x,a,b,c\}$.
Set every terminal-pair precedence relation to $\gtrdot$, except
\[
    x\lessdot a,
    \qquad
    x\lessdot c,
\]
and use the usual boundary relations.
For $L=\{xa\}$, consider two complete deterministic OPTA with
states $\{p,q,r\}$, final state $r$, and the following horizontal
output functions:
\[
\begin{array}{c|cc}
    \text{horizontal word} & \mathcal A_1 & \mathcal A_2\\
    \hline
    a & p & p\\
    b & p & q\\
    c & q & q\\
    xp & r & r\\
    \text{every other word} & q & q
\end{array}
\]
Their unrestricted accepted tree languages are
\[
    \Tree(\mathcal A_1)
      =\{\bullet(x,\bullet(a)),\bullet(x,\bullet(b))\},
    \qquad
    \Tree(\mathcal A_2)
      =\{\bullet(x,\bullet(a))\}.
\]
However,
\[
    T_M(xa)=\bullet(x,\bullet(a)),
    \qquad
    T_M(xb)=\bullet(\bullet(x),b).
\]
Thus, the extra tree accepted by $\mathcal A_1$ is invalid for $M$,
and both OPTA recognize the same word language $L$.

Both require three vertical states among deterministic devices
having a run on every valid input.
Indeed, $\bullet(a)$ and $\bullet(c)$ must reach distinct
nonfinal states: the context $\bullet(x,\Box)$ gives valid
trees for $xa$ and $xc$, and distinguishes them.
An accepting state for $xa$ is distinct from both.
Nevertheless, the two displayed automata are not isomorphic:
$\mathcal A_1$ assigns the same state to $\bullet(a)$ and
$\bullet(b)$, while $\mathcal A_2$ does not.
Hence, minimum recognition on valid inputs alone need not be
canonical, even for an acyclic matrix.
\end{example}

\subsection{The canonical exact OPTA}
\label{sec:canonical-exact-opta}

For a complete deterministic OPTA $\mathcal A$, denote by
$e_{\mathcal A}(t)$ the unique state computed at the root of $t$.
The automaton is reduced if each of its states equals
$e_{\mathcal A}(t)$ for some $t\in\mathcal T_\Sigma^\bullet$.

\begin{lemma}
\label{lem:opta-exact-existence}
If $L$ is recognized by an OPTA over $(\Sigma,M)$, then
$T_M(L)$ is a regular unranked tree language and has an
effectively constructible reduced complete deterministic
exact OPTA.
\end{lemma}

\begin{proof}
Choose an equivalent C-OPG $G$ with
$\OPM(G) \subseteq M$.
Its derivation trees, with nonterminal labels erased, form a
regular unranked tree language: each nonterminal supplies one
state, and the finite union of its production expressions supplies
the corresponding regular horizontal language.
Every accepted derivation tree has underlying tree $T_M(w)$
for its yield $w$, and every $w\in L$ has such a derivation.
Hence, this tree language is exactly $T_M(L)$.

Apply bottom-up determinization and completion, then remove
unreachable states.
These operations are effective for regular horizontal languages.
In particular, the subset construction retains all possible
nonterminal states at each node; its horizontal transition
languages are regular by closure under finite Boolean operations
and finite letter substitutions.
\end{proof}

\begin{definition}
\label{def:opta-syntactic-congruence}
For $s,t\in\mathcal T_\Sigma^\bullet$, define
\[
    s\equiv_{M,L}t
    \quad\Longleftrightarrow\quad
    \forall C\in\mathcal C_\Sigma^\bullet\;
    \bigl(C[s]\in T_M(L)
          \Longleftrightarrow C[t]\in T_M(L)\bigr).
\]
Write $[t]$ for the equivalence class of $t$ and define
\[
    Q_{M,L}=\mathcal T_\Sigma^\bullet/{\equiv_{M,L}}.
\]
\end{definition}

\begin{lemma}
\label{lem:opta-finite-congruence}
The equivalence $\equiv_{M,L}$ has finite index.
Moreover, replacing an internal-rooted child by an equivalent
tree preserves the equivalence class of its parent tree.
\end{lemma}

\begin{proof}
Let $\mathcal D$ be a reduced complete deterministic exact
recognizer with state set $Q_{\mathcal D}$.
If $e_{\mathcal D}(s)=e_{\mathcal D}(t)$, then every context
computes the same root state on $s$ and $t$, and hence
$s\equiv_{M,L}t$.
Therefore, the index of $\equiv_{M,L}$ is at most $|Q_{\mathcal D}|$.

For substitution, fix all other children and any outer context.
The resulting expression is itself a one-hole context.
The defining equivalence therefore permits the replacement.
Replacing several children one at a time gives the corresponding
simultaneous-substitution property.
\end{proof}

For a horizontal word
$\alpha\in(\Sigma\uplus Q_{M,L})^+$, replace each state symbol
$[t]$ by an arbitrary representative tree and each terminal
symbol by its leaf, and place a $\bullet$-root above the resulting
sequence.
Denote the resulting tree by $\rep(\alpha)$.
Lemma~\ref{lem:opta-finite-congruence} shows that its equivalence
class is independent of the chosen representatives.
Hence, the function
\[
    \mu_{M,L}: (\Sigma \uplus Q_{M,L})^+ \to Q_{M,L},
    \qquad
    \mu_{M,L}(\alpha)=[\rep(\alpha)]
\]
is well defined.

\begin{definition}[Canonical exact OPTA]
\label{def:opta-canonical-minimum}
The \emph{canonical exact OPTA} for $(M,L)$ is
\[
    \mathcal A^{\min}_{M,L}
      =\bigl(\Sigma,M,Q_{M,L},F_{M,L},
             (H_c^{\min})_{c\in Q_{M,L}}\bigr),
\]
where
\[
    F_{M,L}=\{[t]\mid t\in T_M(L)\},
    \qquad
    H_c^{\min}
       =\{\alpha\in(\Sigma\uplus Q_{M,L})^+
                    \mid \mu_{M,L}(\alpha)=c\}.
\]
\end{definition}

For complete deterministic OPTA $\mathcal A$ and $\mathcal A'$ over
$(\Sigma,M)$, write $\mu_{\mathcal A}$ and $\mu_{\mathcal A'}$ for
their horizontal output functions.
A \emph{homomorphism} from $\mathcal A$ to $\mathcal A'$ is a map
$\varphi:Q\to Q'$ whose extension $\bar\varphi$ to horizontal words
fixes terminals and satisfies
\[
    \varphi(\mu_{\mathcal A}(\alpha))
      =\mu_{\mathcal A'}(\bar\varphi(\alpha)),
    \qquad
    q\in F\ \Longleftrightarrow\ \varphi(q)\in F',
\]
for every $\alpha\in(\Sigma\uplus Q)^+$ and $q\in Q$.

\begin{theorem}
\label{thm:canonical-minimal-opta}
For every OPTA language $L$ over an OPM $M$:
\begin{enumerate}
    \item $\mathcal A^{\min}_{M,L}$ is a finite reduced
          normalized complete deterministic exact OPTA;
    \item every reduced normalized complete deterministic
          exact OPTA for $(M,L)$ admits a unique surjective
          homomorphism onto $\mathcal A^{\min}_{M,L}$;
    \item $\mathcal A^{\min}_{M,L}$ has the minimum number of
          vertical states among such exact recognizers, and
          every minimum recognizer is isomorphic to it.
\end{enumerate}
Isomorphism identifies horizontal languages after renaming
vertical states; it does not identify arbitrary syntactic
representations of these regular languages.
\end{theorem}

\begin{proof}
Finiteness of $Q_{M,L}$ follows from
Lemma~\ref{lem:opta-finite-congruence}.
We first verify that the horizontal languages are regular.

Let
\[
    \mathcal D=(\Sigma,M,P,F_{\mathcal D},(G_p)_{p\in P})
\]
be a reduced normalized complete deterministic exact recognizer.
Define
\[
    \pi:P\longrightarrow Q_{M,L},
    \qquad
    \pi(e_{\mathcal D}(t))=[t].
\]
This is well defined by the first part of the proof of
Lemma~\ref{lem:opta-finite-congruence}, and it is surjective.
Extend $\pi$ to a letter-to-letter homomorphism
\[
    h:(\Sigma\uplus P)^*\longrightarrow
       (\Sigma\uplus Q_{M,L})^*
\]
that fixes each terminal.
Then, for every $c\in Q_{M,L}$,
\begin{equation}
\label{eq:opta-horizontal-quotient}
    H_c^{\min}
      =
    h\left(\bigcup_{\substack{p\in P\\\pi(p)=c}}G_p\right).
\end{equation}
Indeed, every horizontal word over $P$ can be realized by
choosing representative trees for its state symbols.
Its image under $h$ computes the class of that same parent tree.
Conversely, every class symbol has a representative among
the reachable states of $\mathcal D$.
Equation~\eqref{eq:opta-horizontal-quotient} proves regularity.

The languages $H_c^{\min}$ partition all horizontal
words, so the automaton is normalized, complete, and deterministic.
Induction on trees gives
\[
    e_{\mathcal A^{\min}_{M,L}}(t)=[t].
\]
It follows both that every state is reachable and that the
accepted tree language is exactly $T_M(L)$.

For any reduced complete deterministic exact recognizer $\mathcal D$,
the same map
$\pi$ preserves transitions:
if $\mu_{\mathcal D}(\alpha)$ denotes the unique state selected
by the horizontal word $\alpha$, then
\[
    \pi(\mu_{\mathcal D}(\alpha))
       =\mu_{M,L}(h(\alpha)).
\]
It also preserves and reflects finality:
\[
    p\in F_{\mathcal D}
    \quad\Longleftrightarrow\quad
    \pi(p)\in F_{M,L}.
\]
Hence, $\pi$ is a surjective automaton homomorphism.
Any homomorphism must send the value of each tree $t$ to $[t]$,
so it is uniquely determined because $\mathcal D$ is reduced.

Hence, every reduced complete deterministic exact recognizer has at least
$|Q_{M,L}|$ vertical states.
If it has exactly that many, $\pi$ is bijective and therefore
an isomorphism.

The canonical recognizer is effectively constructible.
Obtain $\mathcal D$ by Lemma~\ref{lem:opta-exact-existence}.
A product of deterministic word automata for the languages $G_p$
gives a finite Moore machine computing $\mu_{\mathcal D}$ on
nonempty words.
Its reachable prefix states and its finite monoid of suffix
transformations allow us to enumerate all maps
\[
    g_{\alpha,\beta}:P\to P,
    \qquad
    g_{\alpha,\beta}(p)=\mu_{\mathcal D}(\alpha p\beta),
    \qquad
    \alpha,\beta\in(\Sigma\uplus P)^*.
\]
Since $\mathcal D$ is reduced, each such map is induced by an
elementary tree context. Conversely, every tree context decomposes
along the path from its hole to its root into elementary contexts.
Closing these maps under composition and adding the identity
therefore computes the finite set $V_{\mathcal D}$ of all context
transformations on $P$.
Now compute
\[
    p\sim q
    \quad\Longleftrightarrow\quad
    \forall v\in V_{\mathcal D}\;
    \bigl(v(p)\in F_{\mathcal D}
          \Longleftrightarrow v(q)\in F_{\mathcal D}\bigr).
\]
By exactness and reducedness, $p\sim q$ holds exactly when
$\pi(p)=\pi(q)$.
Hence, $P/{\sim}$ effectively represents $Q_{M,L}$, and
Equation~\eqref{eq:opta-horizontal-quotient} constructs its regular
horizontal languages.
\end{proof}

This is the tree-context Myhill--Nerode construction specialized
to the OPTA convention of terminal leaves and a single internal
label.

\subsection{The ranked presentation for acyclic matrices}
\label{sec:ranked-presentation}

For the aperiodicity analysis of the next section, 
we also need a ranked presentation
of the exact tree language. In this subsection, assume that $M$
is acyclic.   
Define 
$B = 2|\Sigma|+1$,
$\Gamma_M = \Sigma \cup \{\bullet_k \mid 1\leq k \leq B\}$,
where terminals have rank zero and $\bullet_k$ has rank $k$.
Call $T_{\Gamma_M}$ the set of all trees over this ranked
alphabet. Relabeling each $\bullet$-node of arity $k$ by
$\bullet_k$ identifies the trees in $\mathcal T_\Sigma^\bullet$
of arity at most $B$ with $T_{\Gamma_M}\setminus\Sigma$;
here $\Sigma$ denotes the set of one-leaf terminal trees.
We use this identification implicitly and retain the notation
$T_M(L)$ for the ranked image. Since every valid OP tree has
arity at most $B$, we have $T_M(L)\subseteq T_{\Gamma_M}$;
$T_{\Gamma_M}$ also contains invalid trees and
single terminal leaves.

We apply the standard Myhill-Nerode construction for ranked
tree languages~\cite[Section~1.5]{TATA} to the language $T_M(L)$.
The resulting automaton recognizes precisely $T_M(L)$ and
therefore rejects every tree outside the valid OP-tree universe.

\begin{definition}[Canonical minimal ranked tree automaton]
\label{def:canonical-exact-opta}
A \emph{tree context} is a ranked tree with exactly one hole,
into which a single tree can be substituted.
For $s,t\in T_{\Gamma_M}$, define
\[
    s\equiv^{\mathrm{rk}}_{M,L}t
    \quad
    \Longleftrightarrow
    \quad
    \forall C\;
    \bigl(C[s] \in T_M(L) \Longleftrightarrow C[t] \in T_M(L)\bigr).
\]
The canonical minimal recognizer $\mathcal B_{M,L}$ has state set
$\widehat Q_{M,L}=T_{\Gamma_M}/{\equiv^{\mathrm{rk}}_{M,L}}$.

Writing $[t]_{\mathrm{rk}}$ for the class of $t$, its transitions are
\[
    \Delta_a=[a]_{\mathrm{rk}},
    \qquad
    \Delta_{\bullet_k}([t_1]_{\mathrm{rk}},\ldots,[t_k]_{\mathrm{rk}})
      =[\bullet_k(t_1,\ldots,t_k)]_{\mathrm{rk}},
\]
and its final states are $\{[t]_{\mathrm{rk}}\mid t\in T_M(L)\}$.
\end{definition}

By Lemma~\ref{lem:opta-exact-existence} and the arity bound,
$T_M(L)$ is a regular ranked tree language. The equivalence
therefore has finite index, and its quotient is the reduced
minimal complete deterministic ranked tree automaton recognizing
exactly $T_M(L)$.
In particular, it rejects trees that are not $M$-compatible.
As with the unranked construction, minimality concerns exact
recognition, not behavior specified only on $M$-compatible inputs.

The automata $\mathcal A^{\min}_{M,L}$ and $\mathcal B_{M,L}$ use
different state-counting conventions. The unranked congruence
$\equiv_{M,L}$ has domain $\mathcal T_\Sigma^\bullet$ and uses
the contexts in $\mathcal C_\Sigma^\bullet$; its quotient counts
only states assigned to internal-rooted trees.
The ranked congruence $\equiv^{\mathrm{rk}}_{M,L}$ instead uses
ranked contexts on all of $T_{\Gamma_M}$, including terminal
leaves, so $\mathcal B_{M,L}$ also evaluates leaves in its state set.
The original OPTA continues to treat terminal children by their labels.
Moreover, the unranked exact recognizer must reject every tree
containing a node with more than $B$ children, whereas such trees
lie outside $T_{\Gamma_M}$.
Section~\ref{sec:aperiodicity} uses $\mathcal B_{M,L}$ for the
acyclic aperiodicity characterization.

\section{Tree-context aperiodicity for acyclic matrices}
\label{sec:aperiodicity}

Section~\ref{sec:minimality} supplies canonical recognizers of the
exact OP-tree language. Following the context-aperiodicity framework
of~\cite{DBLP:conf/cai/EsikI07}, we study
the transformations induced by tree contexts on their states.

Throughout this section, let $M$ be acyclic and let
$L\subseteq L(M)$ be an OPTA language. We use the arity bound
$B$, ranked alphabet $\Gamma_M$, and canonical minimal ranked
tree automaton $\mathcal B_{M,L}$ introduced in
Section~\ref{sec:ranked-presentation}.

\begin{definition}[Context aperiodicity]
\label{def:opta-context-aperiodicity}
Let $\mathcal B$ be a reduced complete deterministic ranked
tree automaton with state set $\widehat Q$.
Every tree context $C$ induces a transformation
\[
    \tau_C : \widehat Q \to \widehat Q.
\]
For the canonical recognizer, this transformation is
$\tau_C([t]_{\mathrm{rk}})=[C[t]]_{\mathrm{rk}}$.
The \emph{context transition monoid} is
\[
    V(\mathcal B)=\{\tau_C\mid C\text{ is a tree context}\},
\]
with composition and identity induced by context substitution.

The automaton is \emph{aperiodic} if there exists $N\geq 1$ such that
\[
    \tau^N=\tau^{N+1}
    \qquad\text{for every }\tau\in V(\mathcal B).
\]
We call the canonical OPTA presentation of $L$ aperiodic when
$\mathcal B_{M,L}$ has this property.
\end{definition}

We recall the standard counter-free characterization of context
aperiodicity~\cite{DBLP:conf/cai/EsikI07}.

\begin{theorem}[Counter-free characterization]
\label{prop:opta-counterfree}
For an automaton $\mathcal B$ as above, aperiodicity is
equivalent to
\[
    \tau_C^m(q)=q
    \quad\Longrightarrow\quad
    \tau_C(q)=q
\]
for every context $C$, state $q\in\widehat Q$, and integer $m\geq 1$.
If these conditions hold, then
\[
    \tau^{|\widehat Q|}
    = \tau^{|\widehat Q|+1}
    \qquad\text{for every }\tau\in V(\mathcal B).
\]
Hence, $|\widehat Q|$ is an exponent witnessing aperiodicity.
\end{theorem}

\begin{proof}
A transformation of a finite set has eventually constant powers
exactly when every cycle in its functional graph has length one.
In that case every state reaches a fixed point within $|\widehat Q|$ steps.
This observation can be applied to every transformation in $V(\mathcal B)$.
\end{proof}

Let $\lp$ and $\rp$ be two fresh parenthesis symbols.
The \emph{parenthesis encoding} is
\[
    \enc(a)=a,
    \qquad
    \enc(\bullet_k(t_1,\ldots,t_k))
      =\lp\enc(t_1)\cdots
        \enc(t_k)\rp.
\]
Set $\lp L \rp=\enc(T_M(L))$.
The encoding is injective and, on valid OP trees, agrees with
the canonical parenthesization of~\cite{MPC23}: each internal node
contributes the pair of parentheses surrounding one production
application in an equivalent OP grammar.
We use the non-counting condition of
\cite[Definitions~2.11 and~2.14]{MPC23}:
there exists $n_0\geq 2$ such that, for all $x,u,z,v,y$ with $z$
and $uzv$ well-parenthesized,
\begin{equation}
\label{eq:balanced-nc}
    xu^{n_0}zv^{n_0}y\in \lp L \rp
    \quad\Longleftrightarrow\quad
    xu^{n_0+m}zv^{n_0+m}y\in \lp L \rp
    \qquad(m\geq 0).
\end{equation}

The next lemma accounts for the fact that a balanced factor can
encode a forest, rather than a single subtree.
For this auxiliary argument, a \emph{forest} \cite{BojanczykWalukiewicz2008}
is a finite ordered
sequence of trees built from terminal leaves $a\in\Sigma$ and
$\bullet$-nodes of arbitrary arity, including zero.
We identify a tree with its singleton forest, denote the empty
forest by $0$, and write $+$ for concatenation.
The notation $\bullet(f)$ denotes the tree whose root has label
$\bullet$ and whose ordered sequence of children is $f$.

A \emph{forest context} contains exactly one hole $\Box$ in a
sequence of siblings, possibly at the top level. Formally, such
contexts are generated by
\[
    P ::= \Box \mid f+P+g \mid \bullet(P),
\]
where $f$ and $g$ range over forests.
Substitution $P[f]$ replaces the hole by the entire sequence $f$;
substitution of $0$ deletes the hole. For example, for
$P=\bullet(a+\Box+b)$,
\[
    P[c+d]=\bullet(a+c+d+b),\qquad P[0]=\bullet(a+b).
\]
Composition and powers are defined by substitution as for tree
contexts: $P^0=\Box$ and $P^{n+1}=P[P^n]$.

The parenthesis encoding extends to these forests by
\[
    \enc(0)=\varepsilon,\qquad
    \enc(f+g)=\enc(f)\enc(g),\qquad
    \enc(\bullet(f))=\lp\enc(f)\rp.
\]
We view $K\subseteq T_{\Gamma_M}$ as a forest language containing
only its single-tree members, identifying each $\bullet_k$ with
$\bullet$ of arity $k$.

\begin{lemma}
\label{lem:bounded-forest-pumping}
Let $K\subseteq T_{\Gamma_M}$.
Suppose that for some $N\geq 1$, all ranked tree contexts $C,D$,
and all $t\in T_{\Gamma_M}$,
\begin{equation}
\label{eq:tree-pumping}
    D[C^N[t]]\in K
    \quad\Longleftrightarrow\quad
    D[C^{N+1}[t]]\in K.
\end{equation}
Then, with $n_0=\max\{B+1,N+1\}$, for all forest contexts $P,D$,
all forests $f$, and all $n\geq n_0$,
\begin{equation}
\label{eq:forest-pumping}
    D[P^n[f]]\in K
    \quad\Longleftrightarrow\quad
    D[P^{n+1}[f]]\in K.
\end{equation}
\end{lemma}

\begin{proof}
Equation~\eqref{eq:tree-pumping} also holds with any exponent
$n\geq N$, by replacing $t$ with $C^{n-N}[t]$.

First, suppose the hole of $P$ has no parent.
Then
\(
    P[f]=h_\ell+f+h_r
\)
for fixed forests $h_\ell,h_r$.
If both are empty, $P$ is the identity.
Otherwise $P^n[f]$ has more than $B$ roots for $n>B$.
Substitution into $D$ therefore produces either several top-level
trees or a node with more than $B$ children.
In either case membership in $K$ is impossible.

Then, suppose the hole has a parent.
There are fixed forests $h_\ell,h_r$ and a forest
context $U$ that yields a single tree under every forest substitution,
such that
\(
    P[f]=h_\ell+U[f]+h_r.
\)
Define a one-tree context
\(
    E[t]=U[h_\ell+t+h_r]
\)
and put $t_*=U[f]$.
A direct induction gives
\(
    P^n[f]=h_\ell+E^{n-1}[t_*]+h_r
\), $n\geq 1$.
With $D'[t]=D[h_\ell+t+h_r]$, we obtain
\(
    D[P^n[f]]=D'[E^{n-1}[t_*]].
\)

If $t_*$ is not a tree of $T_{\Gamma_M}$, or $E$ or $D'$ is not
a ranked tree context over $\Gamma_M$, the resulting forest
cannot belong to $K$ for any $n\geq 2$:
it contains a forbidden arity (including zero) or has more than
one root. Such a defect persists under substitution of a single
tree; for $n\geq 2$, the expression contains both $t_*$ and at
least one copy of $E$.
Otherwise, apply~\eqref{eq:tree-pumping} with context $E$,
outer context $D'$, and exponent $n-1\geq N$.
This proves~\eqref{eq:forest-pumping}.
\end{proof}

\begin{theorem}[Aperiodicity and non-counting for acyclic OPTA]
\label{thm:opta-aperiodicity}
Let $M$ be acyclic, and let $L$ be
recognized by an OPTA over $(\Sigma,M)$.
The following conditions are equivalent:
\begin{enumerate}
    \item $\lp L \rp$ satisfies the balanced non-counting
          condition~\eqref{eq:balanced-nc}.
    \item The canonical minimal ranked tree automaton $\mathcal B_{M,L}$
          is aperiodic.
    \item There exists an aperiodic reduced complete deterministic
          ranked tree automaton recognizing exactly $T_M(L)$.
\end{enumerate}
\end{theorem}

\begin{proof}
\emph{(1) implies (2).}
Let $n_0$ witness~\eqref{eq:balanced-nc}.
For arbitrary tree contexts $C,D$ and tree $t$, write
\[
    \enc(C[t])=u\enc(t)v,
    \qquad
    \enc(D[t])=x\enc(t)y.
\]
Here the strings $u,v$ depend only on $C$, and $x,y$ only on $D$.
Putting $z=\enc(t)$ gives
\[
    \enc(D[C^n[t]])=xu^nzv^ny.
\]
Both $z$ and $uzv$ are well-parenthesized.
The non-counting condition therefore yields
\[
    D[C^{n_0}[t]]\in T_M(L)
    \quad\Longleftrightarrow\quad
    D[C^{n_0+1}[t]]\in T_M(L).
\]
Since this holds for every outer context $D$,
\[
    C^{n_0}[t]\equiv^{\mathrm{rk}}_{M,L}C^{n_0+1}[t].
\]
Every state of the canonical recognizer is represented by a tree,
so $\tau_C^{n_0}=\tau_C^{n_0+1}$ for every $C$.

\emph{(2) implies (3).}
Take the canonical recognizer itself.

\emph{(3) implies (1).}
Let $\mathcal B$ recognize exactly $T_M(L)$, with aperiodicity
exponent $N$.
Evaluation in $\mathcal B$ immediately gives
\eqref{eq:tree-pumping}.
Lemma~\ref{lem:bounded-forest-pumping} gives
\eqref{eq:forest-pumping} with uniform threshold
$n_0=\max\{B+1,N+1\}$.

Any well-parenthesized word $z$ encodes a forest $f$.
A well-parenthesized factor of a well-parenthesized word encodes
a consecutive sequence of siblings at some depth, possibly at the
top level. Replacing that sequence by a hole gives a forest context;
an empty factor corresponds to an insertion position.
If $uzv$ is also well-parenthesized, then $u\,\Box\,v$ encodes
a forest context $P$, and
\[
    \enc(P^n[f])=u^nzv^n.
\]
If $xu^nzv^ny$ is well-parenthesized for some $n$, then
$x\,\Box\,y$ similarly defines an outer forest context $D$.
Replacing its balanced middle factor by any other forest encoding
preserves well-parenthesization, so the same $D$ works for every
exponent. If there is no such $n$, all the relevant words lie outside
$\lp L \rp$.
Hence,~\eqref{eq:forest-pumping} gives equality of membership at
every pair of consecutive exponents at least $n_0$.
Iterating these equivalences proves~\eqref{eq:balanced-nc}
for every $m\geq 0$.
\end{proof}

Note that even when every horizontal transition language $H_q$ is finite,
iteration of a tree context can cycle through distinct states.
Therefore, aperiodicity of the individual word languages $H_q$ does
not characterize aperiodicity of the OPTA.

\section{Conclusions}

\begin{figure}[htbp]
\centering
\resizebox{\linewidth}{!}{%
\begin{tikzpicture}[
    x=1cm, y=1cm,
    line width=0.65pt,
    >={Latex[length=2.7mm,width=1.9mm]},
    outer/.style={draw,minimum width=4cm,minimum height=3.4cm,
                  inner sep=0pt,outer sep=0pt},
    inner/.style={draw,minimum width=2.7cm,minimum height=1.2cm,
                  inner sep=0pt,outer sep=0pt,font=\Large},
    title/.style={font=\LARGE,inner sep=0pt},
    ref/.style={font=\normalsize,inner sep=2pt},
    edge/.style={->},
    new edge/.style={edge,draw=blue},
    new ref/.style={ref,text=blue}
]

\node[outer] (opta) at (13,3.3) {};
\node[outer] (mso)  at (8,8.3) {};
\node[outer] (ope)  at (18,8.3) {};
\node[title] at (13,4.25) {$\mathrm{OPTA}$};
\node[title] at (8,9.25)  {$\mathrm{MSO}$};
\node[title] at (18,9.25) {$\mathrm{OPE}$};

\node[inner] (ap) at (13,2.7) {$\mathrm{OPTA}_{\mathrm{AP}}$};
\node[inner] (fo) at (8,7.7) {$\mathrm{FO}$};
\node[inner] (potl) at (3,7.7) {$\mathrm{POTL}$};
\node[inner] (sf) at (18,7.7) {$\mathrm{OPE}_{\mathrm{SF}}$};

\node[outer] (opsc) at (3,3.3) {};
\node[outer] (opa)  at (8,-1.7) {};
\node[outer] (opg)  at (18,-1.7) {};
\node[title] at (3,3.3)  {$\mathrm{OPSC}$};
\node[title] at (8,-1.7)  {$\mathrm{OPA}$};
\node[title] at (18,-2.65) {$\mathrm{OPG}$};
\node[inner] (nc) at (18,-1.0) {$\mathrm{OPG}_{\mathrm{AP}}$};

\draw[edge] ([yshift=2.5mm]potl.east) --
    ([yshift=2.5mm]fo.west);
\draw[edge] ([yshift=-2.5mm]fo.west) --
    ([yshift=-2.5mm]potl.east);
\node[ref] at (5.5,7.7) {$4$};

\draw[edge] ([yshift=4mm]opsc.south east) -- ([xshift=4mm]opa.north west);
\draw[edge] ([yshift=-4mm]opa.north west) -- ([xshift=-4mm]opsc.south east);
\node[ref] at (5.5,0.8) {$3$};

\draw[edge] (7.65,6.6) -- (7.65,0.0);
\draw[edge] (8.15,0.0) -- (8.15,6.6);
\node[ref] at (7.90,3.3) {$1$};

\draw[edge] (10.0,-1.50) -- (16.0,-1.50);
\draw[edge] (16.0,-1.90) -- (10.0,-1.90);
\node[ref] at (13,-1.70) {$1$};

\draw[edge] (10,9.25) -- node[ref,above] {$2$} (16,9.25);
\draw[edge] ([yshift=-2.5mm]sf.west) -- ([yshift=-2.5mm]fo.east);
\node[ref] at (13,7.70) {$2$};

\draw[edge] (19.5,6.6) -- node[ref,right] {$2$} (19.5,0.0);
\draw[edge] (sf.south) -- node[ref,right] {$2$} (nc.north);

\draw[new edge,<->] (ap.south) -- node[new ref,above right] {$5$}
    (nc.west);

\draw[new edge] (opta.south west) -- node[new ref,above left] {$5$}
    (opa.north east);
\draw[new edge] (opg.north west) -- node[new ref,above right] {$5$}
    (opta.south east);

\end{tikzpicture}
}
\caption{Overview of the main results for OPLs with acyclic matrices: 
1 stands for \cite{LonatiEtAl2015}, 
2 for \cite{MPC23}, 
3 for \cite{Henzinger23}, 
4 for \cite{CMP21arXiv}, and
5 (shown in blue) for results proved here.
}
\label{fig:overview}
\end{figure}
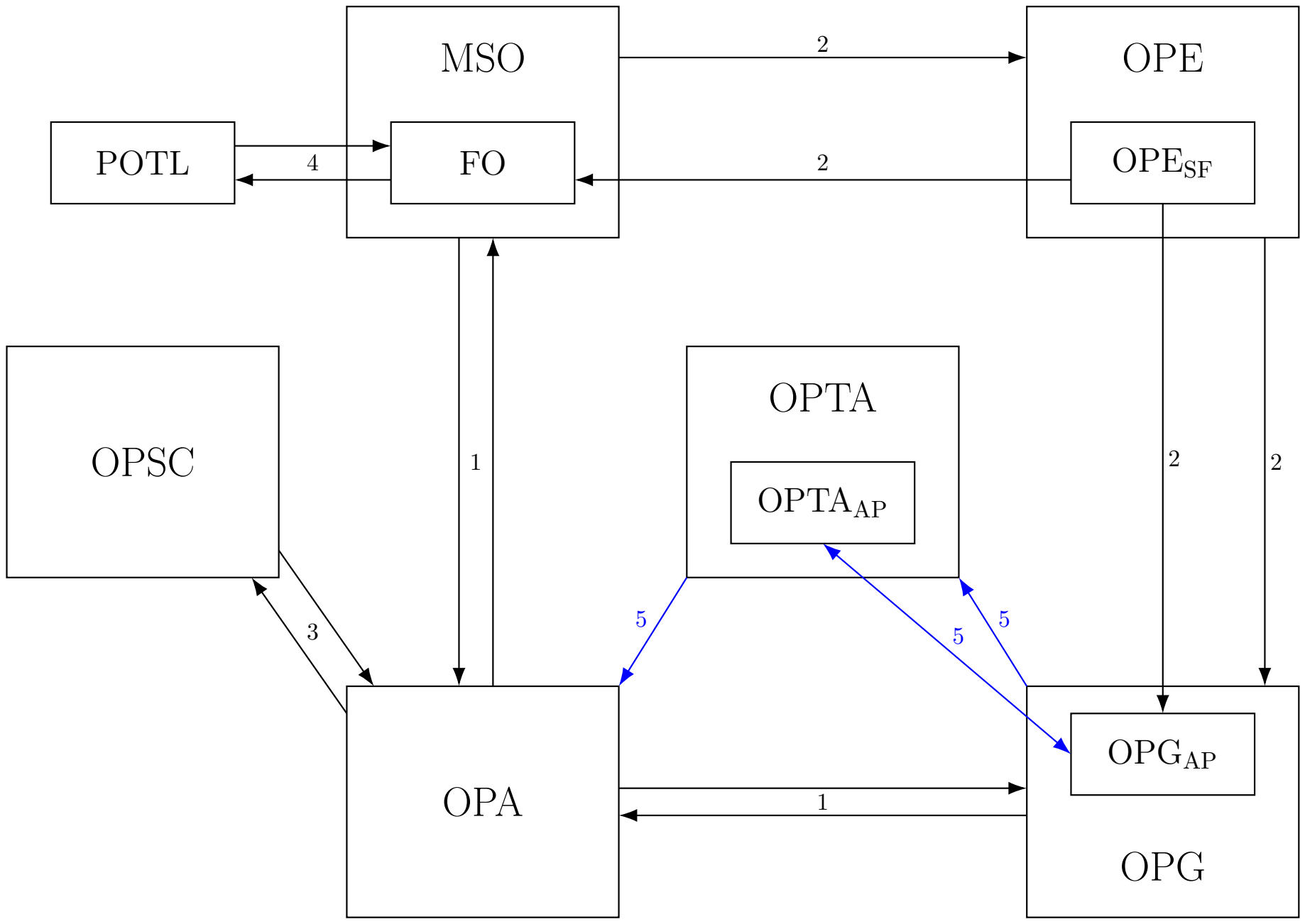

The concept of OPTA offers a useful bridge from the theory of Tree Automata
to the one of Operator Precedence languages that we used to obtain
the following results:

\begin{itemize}
\item We proved that OPTA recognize exactly the OPLs. 
The equivalence holds for arbitrary conflict-free matrices,
including cyclic ones.
\item We constructed an effectively computable canonical complete
deterministic \emph{exact} OPTA, minimal in its number of vertical states.
\item We characterized
balanced non-counting by aperiodicity of the context transition
monoid of the canonical minimal ranked tree recognizer,
in the setting of acyclic matrices.
\end{itemize}

An overview of the updated results is presented in Figure~\ref{fig:overview},
where the boxes represent the languages defined by the following notations:
OPTA$_{\mathrm{AP}}$ denotes languages whose canonical ranked
recognizer is aperiodic;
Monadic Second Order logic for OP is MSO, and FO its first-order fragment;
POTL denotes Precedence-Oriented Temporal Logic;
OP expressions are OPE, and OPE$_{\mathrm{SF}}$ their star-free fragment;
OPG$_{\mathrm{AP}}$ are the aperiodic OP grammars;
OPSC denotes the characterization by the finite-index
OP syntactic congruence of~\cite{Henzinger23}.


\bibliographystyle{alpha}
\bibliography{opbib}

@inproceedings{CristauLoedingThomas2005,
  author = {Julien Cristau and Christof L{\"o}ding and Wolfgang Thomas},
  title = {Deterministic Automata on Unranked Trees},
  booktitle = {Foundations of Semistructured Data},
  series = {Dagstuhl Seminar Proceedings},
  volume = {05061},
  pages = {1--12},
  year = {2005},
  doi = {10.4230/DagSemProc.05061.3}
}

@incollection{BojanczykWalukiewicz2008,
  author    = {Boja{\'n}czyk, Miko{\l}aj and Walukiewicz, Igor},
  title     = {Forest Algebras},
  booktitle = {Logic and Automata: History and Perspectives},
  editor    = {Flum, J{\"o}rg and Gr{\"a}del, Erich and Wilke, Thomas},
  publisher = {Amsterdam University Press},
  series    = {Texts in Logic and Games},
  volume    = {2},
  pages     = {107--132},
  year      = {2008}
}

@article{CMP21arXiv,
  title={A First-Order Complete Temporal Logic for Structured Context-Free Languages},
  author={Michele Chiari and Dino Mandrioli and Matteo Pradella},
  journal   = {Logical Methods in Computer Science},
  year = {2022},
  url = {https://arxiv.org/abs/2105.10740},
  volume       = {18},
  number       = {3},
}

@Article{MPC23,
 author = {Dino Mandrioli and Matteo Pradella and Stefano {Crespi Reghizzi}},
 title = {Aperiodicity, Star-freeness, and First-order Definability of Operator Precedence Languages},
  journal   = {Logical Methods in Computer Science},
  year = {2023},
  Volume = {19}, 
  Number = {4},
  doi = {10.46298/lmcs-19(4:12)2023},
  url = {https://doi.org/10.46298/lmcs-19(4:12)2023},
}

@inproceedings{Henzinger23,
  author       = {Thomas A. Henzinger and
                  Pavol Kebis and
                  Nicolas Mazzocchi and
                  N. Ege Sara{\c{c}}},
  editor       = {Kousha Etessami and
                  Uriel Feige and
                  Gabriele Puppis},
  title        = {Regular Methods for Operator Precedence Languages},
  booktitle    = {50th International Colloquium on Automata, Languages, and Programming,
                  {ICALP} 2023, July 10-14, 2023, Paderborn, Germany},
  series       = {LIPIcs},
  volume       = {261},
  pages        = {129:1--129:20},
  year         = {2023},
  doi          = {10.4230/LIPIcs.ICALP.2023.129},
}

@Misc{TATA,
  author = {Hubert Comon and Max Dauchet and R{\'e}mi Gilleron and Christof L\"oding
and Florent Jacquemard
and Denis Lugiez and Sophie Tison and Marc Tommasi},
  title = {Tree Automata Techniques and Applications},
  howpublished = {https://hal.inria.fr/hal-03367725},
  year = 2008
}

@Article{jacm/AlurM09,
  Title                    = {Adding nesting structure to words},
  Author                   = {Rajeev Alur and Parthasarathy Madhusudan},
  Journal                  = {J. {ACM}},
  Year                     = {2009},
  Number                   = {3},
  Volume                   = {56},

  Bibdate                  = {2009-05-20},
  Bibsource                = {DBLP,
 http://dblp.uni-trier.de/db/journals/jacm/jacm56.html#AlurM09}
}

@Article{BarenghiEtAl2015,
  Title                    = {Parallel Parsing made practical},
  Author                   = {Alessandro Barenghi and Stefano {Crespi Reghizzi} and Dino Mandrioli and Federica Panella and Matteo Pradella},
  journal   = {Sci. Comput. Program.},
    volume    = {112},
    number    = {3},
    pages     = {195--226},
  Year                     = {2015},
  Publisher                = {Elsevier},
  doi = {10.1016/j.scico.2015.09.002},
}

@article{LonatiEtAl2015,
  author    = {Violetta Lonati and
               Dino Mandrioli and
               Federica Panella and
               Matteo Pradella},
  title     = {Operator Precedence Languages: Their Automata-Theoretic and Logic
               Characterization},
  journal   = {{SIAM} J. Comput.},
  volume    = {44},
  number    = {4},
  pages     = {1026--1088},
  year      = {2015},
  doi = {10.1137/140978818},
  bibsource = {dblp computer science bibliography, http://dblp.org}
}

@Article{Crespi-ReghizziM12,
  Title                    = {{Operator Precedence and the Visibly Pushdown Property}},
  Author                   = {Stefano {Crespi Reghizzi} and Dino Mandrioli},
    journal   = {J. Comput. Syst. Sci.},
  Year                     = {2012},
  Number                   = {6},
  Pages                    = {1837--1867},
  Volume                   = {78},
  doi = {10.1016/j.jcss.2011.12.006}
}

@Article{Floyd1963,
  Title                    = {{Syntactic Analysis and Operator Precedence}},
  Author                   = {Robert W. Floyd},
  Journal                  = {J. {ACM}},
  Year                     = {1963},
  Number                   = {3},
  Pages                    = {316--333},
  Volume                   = {10}
}

@INPROCEEDINGS{DBLP:conf/icalp/Mehlhorn80,
AUTHOR = {Kurt Mehlhorn},
TITLE = {Pebbling Mountain Ranges and its Application of {DCFL}-Recognition},
BOOKTITLE = {Automata, languages and programming (ICALP-80)},
YEAR = {1980},
VOLUME = {85},
PAGES = {422--435},
SERIES = {LNCS},
ISBN = {3-540-10003-2},
}

@Book{Salomaa73,
  Title                    = {Formal {L}anguages},
  Author                   = {Arto K. Salomaa},
  Publisher                = {Academic Press},
  Year                     = {1973},

  Address                  = {New York, NY}
}

@article{MP18,
  author    = {Dino Mandrioli and
               Matteo Pradella},
  title     = {Generalizing input-driven languages: Theoretical and practical benefits},
  journal   = {Computer Science Review},
  volume    = {27},
  pages     = {61--87},
  year      = {2018},
  doi       = {10.1016/j.cosrev.2017.12.001},
  bibsource = {dblp computer science bibliography, https://dblp.org}
}

@inproceedings{DBLP:conf/cai/EsikI07,
  author    = {Zolt{\'{a}}n {\'{E}}sik and
               Szabolcs Iv{\'{a}}n},
  title     = {Aperiodicity in Tree Automata},
  booktitle = {Algebraic Informatics, 2nd International Conference, {CAI}},
  pages     = {189--207},
  year      = {2007},
  doi       = {10.1007/978-3-540-75414-5\_12},
}

@article{CMP25,
  author       = {Michele Chiari and
                  Dino Mandrioli and
                  Matteo Pradella},
  title        = {Cyclic operator precedence grammars for parallel parsing},
  journal      = {Inf. Comput.},
  volume       = {307},
  pages        = {105363},
  year         = {2025},
  url          = {https://doi.org/10.1016/j.ic.2025.105363},
  doi          = {10.1016/J.IC.2025.105363},
  bibsource    = {dblp computer science bibliography, https://dblp.org}
}


\end{document}